\documentclass[review]{elsarticle}
\usepackage{booktabs}
\usepackage{txfonts}
\usepackage[usenames, dvipsnames]{color}
\usepackage{lineno,hyperref}
\modulolinenumbers[5]

\journal{Journal of \LaTeX\ Templates}
\newcommand{\beq}{\begin{eqnarray}}
\newcommand{\eeq}{\end{eqnarray}}

\begin{document}

\begin{frontmatter}

\title{Impossibility of Increasing N$\acute{\textrm{e}}$el Temperature in Zigzag Graphene Nanoribbon by Electric Field and Carrier Doping}

\author[mymainaddress]{Teguh Budi Prayitno\corref{mycorrespondingauthor}}

\cortext[mycorrespondingauthor]{Corresponding author}

\address[mymainaddress]{Physics Department, Faculty of Mathematics and Natural Science, Universitas Negeri Jakarta, Kampus A Jl. Rawamangun Muka, Jakarta Timur 13220, Indonesia}

\begin{abstract}


We investigated the dependence of N$\acute{\textrm{e}}$el temperature as a critical temperature on the electric field and hole-electron doping in the antiferromagnetically ordered zigzag graphene nanoribbon. The temperature was calculated by averaging the magnon energy in the Brillouin zone within the mean-field approximation. We employed the generalized Bloch theorem instead of the supercell approach to reduce the computational cost significantly to obtain the magnon spectrum. We showed that the N$\acute{\textrm{e}}$el temperature reduces when increasing both the electric field and the hole-electron doping, thus these treatments will never enhance the N$\acute{\textrm{e}}$el temperature. 
\end{abstract}
\end{frontmatter}


\section{Introduction}
Recently, exploring the electric and magnetic properties in the low-dimensional materials gives significant impacts in the condensed matter subject. It was started from the discovery of graphene as a two-dimensional material composed of carbon atoms by Novoselov $et$ $al.$ \cite{Novoselov1,Novoselov2,Geim}. Besides the abundance of carbon in nature, it was verified that graphene has high electrical and thermal conductivities, exhibiting interesting physical properties. Previous works reported that graphene can be applied well such as for the optoelectronics \cite{Wang, Bonaccorso} or transistor \cite{Schwierz, Liao}. Next, experimental results and theoretical studies of replicas of graphene such as germanene and silicene also show promising materials for the future nanoelectronic devices. Compared to the graphene, germanene and silicine have intrinsic gap \cite{Bayani, Vali} which can be controlled by strain \cite{Yan} or electric field \cite{Ghosal}. From this benefit, germanene and silicine can be more applicable for the logic-based devices such as transistor. Then, performing the experiments or density functional theory (DFT), the future applicable devices based on low-dimensional materials are explored such as for semiconductors \cite{Wang1, Mak} or thermoelectric materials \cite{Lee, Sharma}.     

The main question regarding the low-dimensional materials is related to the critical temperature (Curie or N$\acute{\textrm{e}}$el temperature), at which the magnetism in any materials is lost. In the bulk materials, such as 3$d$ ferromagnetic metals \cite{Uhl, Shall}, Heusler alloys \cite{Enkovaara, Lezaic}, or 3$d$ transition metal oxides \cite{Essenberg}, the critical temperatures are always higher than the room temperature, thus any practical devices based on these materials will operate properly. On the contrary, within the DFT, the critical temperature in the low-dimensional systems, such as 1-T transition metal dihalides monolayer \cite{Kulish, Botana} and most of transition metal dichalcogenides monolayer \cite{He}, are predicted to be lower than the room temperature. So, the magnetism for these low-dimensional materials should be lost at room temperature. 

Regarding the low-dimensional materials, the critical temperature in the zigzag graphene nanoribbons (ZGNR), a one-dimensional structure of graphene, is not thoroughly elucidated. In the previous DFT calculations, Yazyev and Katsnelson \cite{Yazyev1} with the supercell approach stated that the critical temperature in the ZGNR only reaches the room temperature if the order of spin correlation length is only a few nanometers, a subtle feature that is very difficult to realize now. At the same time, Kunstmann $et$ $al.$ \cite{Kunstmann} also claimed that the magnetism in the ZGNR only preserves at a very low temperature. Based on their reports, the magnetism in the ZGNR is only stable below room temperature. As a consequence, any practical devices based on ZGNR will never function well. 

The purpose of this paper is to investigate the influence of the N$\acute{\textrm{e}}$el temperature as a critical temperature of ZGNR with respect to the electric field and hole-electron doping based on the spin-waves excitations within frozen magnon method. The calculation of the N$\acute{\textrm{e}}$el temperature in the antiferromagnetic edge state ZGNR will be performed by the mean-field approximation (MFA) within the generalized Bloch theorem (GBT). The benefit of using the GBT rather than the supercell approach is the efficiency to obtain not only the N$\acute{\textrm{e}}$el temperature but also the spin stiffness through the primitive cell. As reported in the previous paper \cite{Edwards}, in an $sp$-electron system as in the ZGNR, the Stoner excitations may not be neglected. If so, the calculation of spin stiffness will only possible in the low magnon energy close to $\Gamma$ point, thus it is very difficult to realize through the supercell approach.    

To obtain the N$\acute{\textrm{e}}$el temperature, we average the magnon energies for a set of spiral vectors in the Brillouin zone. Here, we exploit the conical spiral instead of the flat spiral to obtain constant magnetic moments during the self-consistent calculation. This approach was successfully employed to estimate critical temperatures in some materials \cite{Halilov, Sasioglu, Sandratskii, Teguh}. We prove that both the electric field and hole-electron doping reduce the N$\acute{\textrm{e}}$el temperature as well as the spin stiffness for all ribbon widths, making the impossibility to reach the room temperature in the ZGNR with these treatments.    

These findings are caused by the small magnetic moments of magnetic carbon atoms at the edges by applying both the electric field and the hole-electron doping. By using the Hubbard approach, Kunstmann $et$ $al.$ \cite{Kunstmann} showed that at the such condition the magnetism in the ZGNR becomes unstable, thus disappearing the magnetic properties at room temperature. So, our results are in good agreement with the former prediction. This means that even though the electric field and hole-electron doping can generate some magnetic properties for spintronic applications, it will not operate well at room temperature.     

\section{Structure Model and Calculation Method}
We applied the GBT within the first-principles calculation as implemented in the OpenMX code \cite{openmx}, a DFT package exploiting the localized basis function \cite{Ozaki} and norm-conserving pseudopotentials \cite{Troullier}, with a 150 Ryd cutoff energy and employed the generalized gradient approximation (GGA) \cite{Perdew} as the exchange-correlation functional for the electron-electron interaction. The implementation of GBT is to express the non-collinear wavefunction as the linear combination of pseudo-atomic orbitals (LCPAOs) by including the spiral wavevector $\mathbf{q}$ 
\beq
\psi_{\nu\mathbf{k}}\left(\mathbf{r}\right)&=&\frac{1}{\sqrt{N}}\left[\sum_{n}^{N}e^{i\left(\mathbf{k}-\frac{\mathbf{q}}{2}\right)\cdot\mathbf{R}_{n}}\sum_{i\alpha}C_{\nu\mathbf{k},i\alpha}^{\uparrow}\phi_{i\alpha}\left(\mathrm{\mathbf{r}-\tau_{i}-\mathbf{R}_{n}}\right)
\left(
\begin{array}{cc}
1\\
0\end{array} 
\right)\right.\nonumber\\
& &\left.+\sum_{n}^{N}e^{i\left(\mathbf{k}+\frac{\mathbf{q}}{2}\right)\cdot\mathbf{R}_{n}}\sum_{i\alpha}C_{\nu\mathbf{k},i\alpha}^{\downarrow}\phi_{i\alpha}\left(\mathrm{\mathbf{r}-\tau_{i}-\mathbf{R}_{n}}\right)\left(
\begin{array}{cc}
0\\
1\end{array}
\right)\right].\label{lcpao}
\eeq
Here, the localized orbital function $\phi_{i\alpha}$ can be generated as many as possible by means of the confinement technique \cite{Ozaki2}.

To evaluate the N$\acute{\textrm{e}}$el temperature and the spin stiffness, we employed the frozen magnon method and mapped the total energy difference in the self-consistent calculation of the spiral magnetic configurations \cite{Sandratskii2}
\beq
\textit{\textbf{M}}_{i}(\mathbf{r}+\mathbf{R}_{i})=M_{i}(\mathbf{r}) \left(
\begin{array}{cc}
\cos\left(\varphi_{0}+\mathbf{q}\cdot \mathbf{R}_{i}\right)\sin\theta_{i}\\
\sin\left(\varphi_{0}+\mathbf{q}\cdot \mathbf{R}_{i}\right)\sin\theta_{i}\\
\cos\theta_{i}\end{array} 
\right) \label{moment}  
\eeq
onto the Heisenberg Hamiltonian model. Within the frozen magnon method, the magnon energy of the ZGNR can be formulated as \cite{Teguh1}
\beq 
\hbar\omega_{\mathbf{q}}=\frac{\mu_{B}}{M}\frac{E(\mathbf{q},\theta)-E(\mathbf{0},\theta)}{\sin^{2}\theta}.\label{magnonf} 
\eeq
Meanwhile, the N$\acute{\textrm{e}}$el temperature within the MFA can be estimated by \cite{Padja}
\beq
k_{\scriptsize{B}}T_{\scriptsize{C}}^{\scriptsize{\textrm{MFA}}}=\frac{M}{6\mu_{B}}\frac{1}{N}\sum_{\mathbf{q}}\hbar\omega_{\mathbf{q}}, \label{MFA}
\eeq
with $N$ denotes the number of $\mathbf{q}$.
   
\begin{figure}[h!]
\vspace{2mm}
\quad\quad\quad\quad\quad\quad\includegraphics[scale=0.5, width =!, height =!]{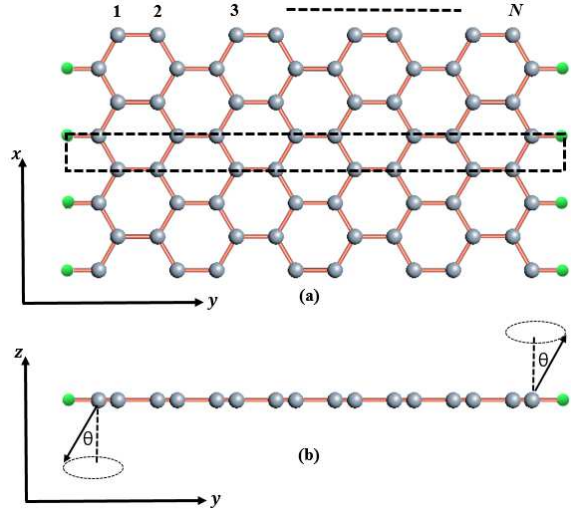}
\vspace{2mm}
\caption{\label{model}(Color online) Top view (a) and side view (b) of $N$-ZGNR antiferromagnetic edge states with the ribbon width $N$. The large and small filled spheres denote carbon and hydrogen atoms, respectively. Meanwhile, the primitive cell is pointed out by a dashed line.} 
\end{figure}

We applied the conical spiral ($\theta=10^{\circ}$) to fix the antiferromagnetically ordered magnetic moments of carbon atoms at the edges, as shown in Fig. \ref{model}. This can be realized by introducing the penalty functional if the magnetic moments start to deviate \cite{Kurz}. For the atomic structure, an experimental lattice of graphite of 2.46 {\AA} as a unit cell in the \emph{x}-axis was applied while the vacuums in the non-periodic cells in the other axes were set to 50 {\AA}. For the basis sets, we used two $s$- and two $p$- orbitals for the carbon atoms, and two $s$- and a $p$- orbitals for the hydrogen atoms. At the same time, the boundary cutoff radii were assigned to 4.0 a.u. and 6.0 a.u. for the carbon and hydrogen atoms, respectively. 

\section{Results and Discussions}
We divide this section into two subsections exploring the influence of electric field and hole-electron doping on the ZGNR. Here, we provide the magnon spectra for $N$-ZGNR ($N=6, 8, 12$) in the Brillouin zone, where $N$ is the ribbon width. First of all, for the non-electric-field and non-doped cases, the magnon energy increases as $N$ increases, thus increasing the N$\acute{\textrm{e}}$el temperature and the spin stiffness. However, since there are flat high energy dispersions around 300 meV, the N$\acute{\textrm{e}}$el temperatures will not reach the room temperature.
 
\subsection{Electric field case}
The application of electric field in the ZGNR is very important to study the magnetic features. When the transverse electric field is applied along the ribbon width, the half-metallic feature is induced \cite{Son, Rudberg, Kan}. Even though this feature is very useful for developing spintronic devices, however, we prove that the ZGNR-based applicable devices cannot operate well at room temperature since the N$\acute{\textrm{e}}$el temperature reduces due to transverse electric field.  
\begin{figure}[h!]
\vspace{4mm}
\quad\quad\includegraphics[scale=0.5, width =!, height =!]{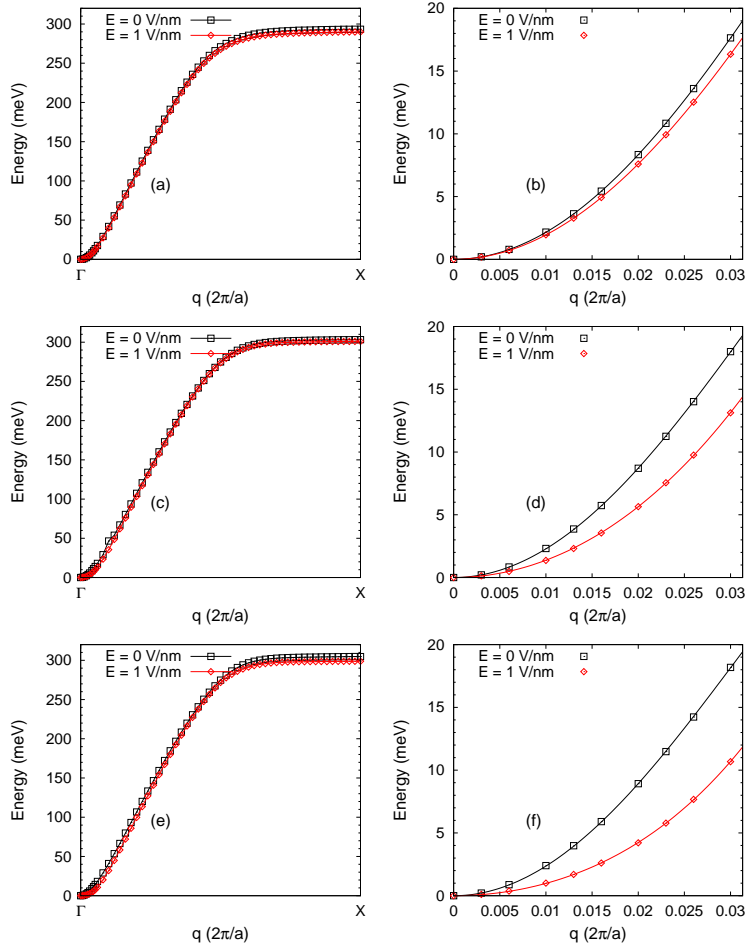}
\vspace{2mm}
\caption{\label{magnon-E}(Color online) Magnon dispersions of ZGNR in the Brillouin zone (a, c, e) and low magnon energies close to $\Gamma$ point (b, d, f) under electric field $E$. The solid lines in (b, d, f) represent the fitting function $\hbar\omega_{q}=D q^{2}(1-\beta q^{2})$. Here, 6-ZGNR, 10-ZGNR, and 12-ZGNR are depicted by figures (a, b), (c, d), and (e, f), respectively.} 
\end{figure}     

Here, we apply the transverse electric field $E$ along $N$ in the \emph{y}-axis and plot the magnon spectra for $E=0$ V/nm and $E=1$ V/nm, as shown in Fig. \ref{magnon-E}. As immediately observed, the applied $E$ reduces all the magnon spectra for each $N$. We also see that there are still flat dispersions for all $N$ in the one-third of the Brillouin zone at the high energies near $\textrm{X}$ point as $E$ increases. Meanwhile, the low energies near $\Gamma$ point also reduce for all $N$ as $E$ increases, reducing the spin stiffness.
\begin{figure}[h!]
\centering
\vspace{-2mm}
\includegraphics[scale=0.5, width =!, height =!]{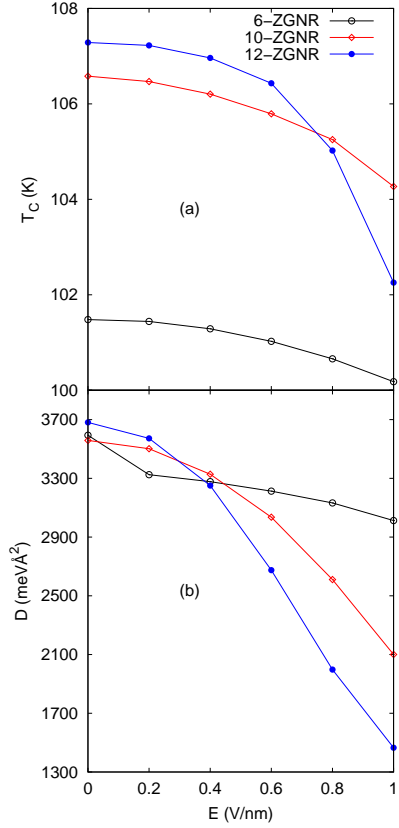}
\vspace{-10mm}
\caption{\label{D-T-E}(Color online) Electric field $E$ dependence of N$\acute{\textrm{e}}$el temperature $T_{C}$ and spin stiffness $D$.} 
\end{figure} 

Based on Figs. \ref{magnon-E}(a), \ref{magnon-E}(c), and \ref{magnon-E}(e), we calculate the N$\acute{\textrm{e}}$el temperature $T_{\scriptsize{C}}$ by averaging all the magnon energies in the Brillouin zone by means of the MFA approach in Eq. (\ref{MFA}). Meantime, the spin stiffness $D$ is evaluated in the low magnon energies near $\Gamma$ point as shown in Figs.  \ref{magnon-E}(b), \ref{magnon-E}(d), and \ref{magnon-E}(f) through the least-square fit $\hbar\omega_{q}=D q^{2}(1-\beta q^{2})$. To view the reduction more clearly, we provide Fig. \ref{D-T-E} to show the reductions of $T_{\scriptsize{C}}$ and $D$ as $E$ increases. Thus, the applied $E$ cannot enhances the $T_{\scriptsize{C}}$ up to the room temperature.
\begin{figure}[h!]
\centering
\vspace{-8mm}
\includegraphics[scale=0.5, width =!, height =!]{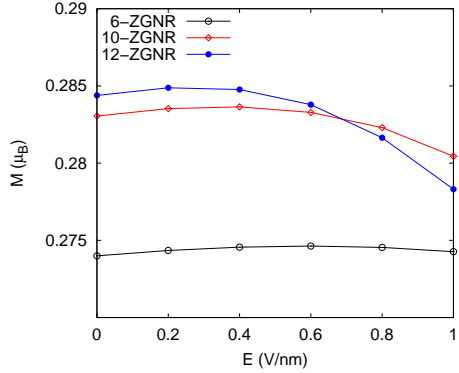}
\vspace{2mm}
\caption{\label{M-E}(Color online) Magnetic moment $M$ as a function of electric field $E$.} 
\end{figure} 

As shown in Fig. \ref{D-T-E}, both the $T_{\scriptsize{C}}$ and $D$ reduce as $N$ increases when $E$ increases. We notice that the reductions of $T_{\scriptsize{C}}$ as well $D$ are more rapid for the large $N$ that those for the small $N$ as $E$ increases. If we consider that the exchange interaction $J_{ij}$ depends only on the distance between two edge carbon atoms, 6-ZGNR should have the largest $J_{ij}$. In this case, the electron from one edge carbon atom hops more easily to the other edge carbon atom in the small $N$ than that in the large $N$. As a consequence, the large $J_{ij}$ may prohibit the rapid reduction for the $T_{\scriptsize{C}}$ and $D$ which is caused by $E$.   

The origin of impossibility of increasing the $T_{\scriptsize{C}}$ is caused by the small magnetic moments of the edge carbon atoms. Our calculation finds the magnetic moment of each edge carbon atom is about 0.3 $\mu_{\textrm{\scriptsize{B}}}$. When $E$ is applied, the magnetic moment generally decreases for all $N$ as shown in Fig. \ref{M-E}, in good agreement with Ref. \cite{Culchac}. In addition, we also see that the reduction of magnetic moment under $E$ for the large $N$ is more rapid than that for the small $N$, the same tendency as in the reduction of $T_{\scriptsize{C}}$ and $D$. According to Kunstmann $et$ $al.$ \cite{Kunstmann}, this small magnetic moment in the ZGNR yields magnetic instability, namely, the magnetism in the ZGNR cannot hold at room temperature. It is also supported by the previous authors who reported the reduction of $D$ under $E$ \cite{Rhim, Teguh3, Teguh4}.  
                 
\subsection{Hole-electron doping case}
It has been reported that the hole-electron doping can induce the magnetic phase transition from ferromagnetic-canted-antiferromagnetic states \cite{Sawada}. So, the implementation will be important to control the magnetic state for the applicable devices. Applying the hole-electron doping can be realized by employing the chemical doping or field effect transistor (FET) doping. In this calculation, we exploit the Fermi level shift (FLS) approach where the system is neutralized by inserting the uniform background charge. Here, we also show that the hole-electron doping also cannot increase the $T_{\scriptsize{C}}$.
\begin{figure}[h!]
\vspace{-2mm}
\quad\quad\includegraphics[scale=0.5, width =!, height =!]{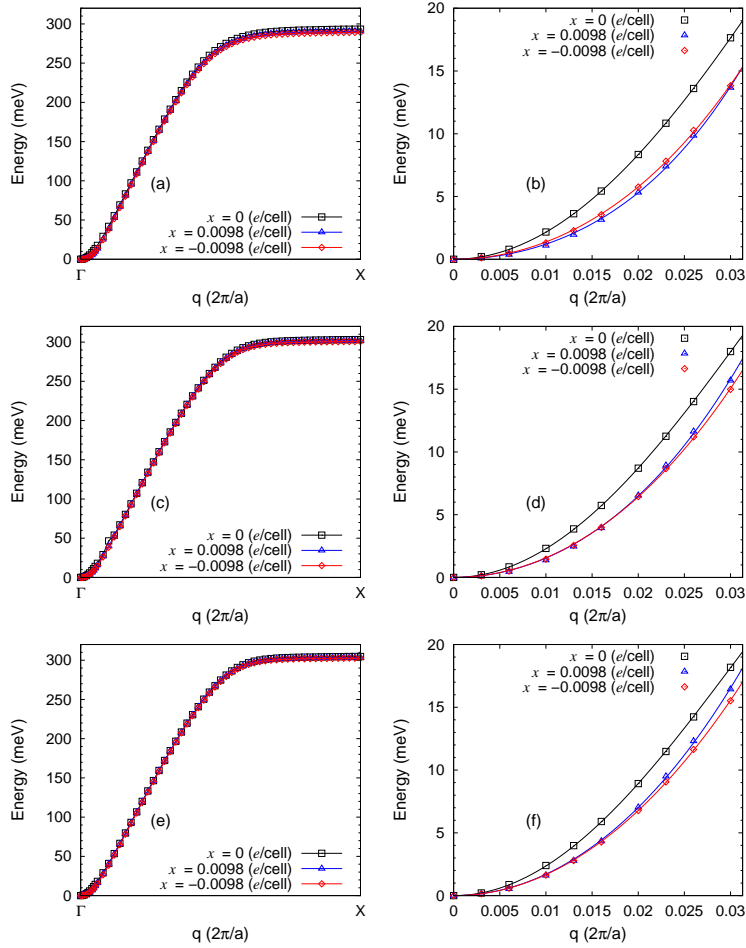}
\vspace{2mm}
\caption{\label{magnon-x}(Color online) Magnon dispersions of ZGNR in the Brillouin zone (a, c, e) and low magnon energies close to $\Gamma$ point (b, d, f) under doping $x$. The solid lines in (b, d, f) represent the fitting function $\hbar\omega_{q}=D q^{2}(1-\beta q^{2})$. Here, 6-ZGNR, 10-ZGNR, and 12-ZGNR are depicted by figures (a, b), (c, d), and (e, f), respectively.} 
\end{figure}     

We plot the magnon spectra via self-consistent calculation as performed in the $E$ case. As shown in Fig. \ref{magnon-x}, we also observe the reduction of magnon spectra for each $N$ as the doping $x$ increases. In addition, the flat dispersions are still observed for all $N$ in the one-third of the Brillouin zone $\textrm{X}$ point as the doping increases. This indicates that the flat dispersions may be the natural feature of magnon dispersion in the ZGNR. We also notice that increasing $x$ will also reduce the $T_{\scriptsize{C}}$ and $D$.  

\begin{figure}[h!]
\centering
\vspace{-8mm}
\includegraphics[scale=0.5, width =!, height =!]{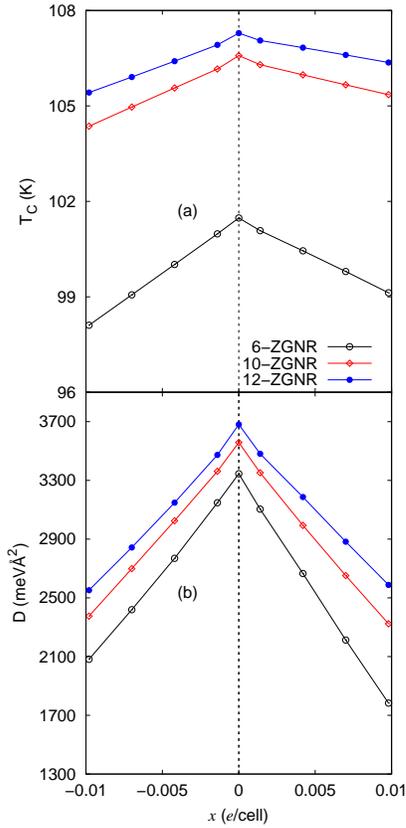}
\vspace{-10mm}
\caption{\label{D-T-x}(Color online) Doping $x$ dependence of N$\acute{\textrm{e}}$el temperature $T_{C}$ and spin stiffness $D$.} 
\end{figure}

\begin{figure}[h!]
\centering
\vspace{-2mm}
\includegraphics[scale=0.5, width =!, height =!]{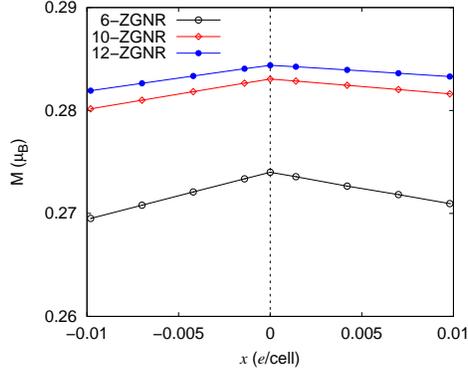}
\vspace{2mm}
\caption{\label{M-x}(Color online) Magnetic moment $M$ as a function of doping $x$.} 
\end{figure}

By applying the same way to calculate the $T_{\scriptsize{C}}$ and $D$ as in the $E$ case, both the $T_{\scriptsize{C}}$ and $D$ incline to reduce as $x$ increases. Figure \ref{D-T-x} shows the reduction of $T_{\scriptsize{C}}$ and $D$ when increasing $x$. When we consider the trends of magnetic moments of edge carbon atoms, we also see the asymmetric reduction of magnetic moment as shown in Fig. \ref{M-x}, similar to Ref. \cite{Teguh1}. This means that taking the doping into account also yields the magnetic instability as in the $E$ case. Unlike the $E$-field case, no rapid reduction of the $T_{\scriptsize{C}}$ and $D$ as $x$ increases, indicating that $x$ does not influence the rapid reduction of the $T_{\scriptsize{C}}$ and $D$ for the large $N$. 

The linear dependence of magnetic moment $M$ on $x$ can be explained as follows. The calculations of magnon energy apply the Heisenberg model in which the total energy difference $\Delta E=E(q)-E(q=0)$ in the self-consistent calculation is mapped onto the Heisenberg Hamiltonian, as stated in Eq. \ref{magnonf}. In this case, $\Delta E$ is proportional to $M$. We then find that $\Delta E$ decreases linearly as the doping increases, the same tendency with Ref. \cite{Sawada}. The reduction of $\Delta E$ leads to a loss of magnetism gradually in ZGNR due to low concentration of doping. So, the reduction should be linear. At the same time, due to proportionality between $\Delta E$ and $M$, $M$ should also reduce linearly as the doping increases. When the doping is sufficiently high, the ZGNR should become non-magnetic. Note that introducing hydrogen passivation at the edge will remove the dangling bond state, thus reducing the magnetic moment of edge carbon atoms, as pointed out by Song $et$ $al.$ \cite{Song}. This means that introducing the doping or electric field reduces the magnetism in the ZGNR.

Regarding the doping case, we give some comments on the possibility to increase the $T_{\scriptsize{C}}$. Since the main problem of small magnon energy is the small magnetic moment of each edge carbon atom, the most possible way is to introduce the metal atom especially with the large magnetic moment. As reported in Ref. \cite{Kunstmann}, the small magnetic moment leads to unstable magnetism that vanishes the magnetism at room temperature. When the metal atom is introduced, it forms a strong bonding between the edge carbon atom and metal atom which transfer the charge from the metal atom to the edge atom, thus increasing the magnetic moment of edge carbon atom. 

For the related experiment, Magda $et$ $al.$ \cite{Magda} grew the ZGNR onto Au(111) substrate by chemical deposition. They justified that this material will be stable at room temperature. In the computational framework, previous authors also showed that the robust magnetism in the ZGNR can be achieved by introducing the metal atoms when considering the spiral density waves \cite{Huang, Liang, Lu}. They found the large scale energy of spiral states that can be observed at room temperature. Since the spiral spin density waves are manifestation of spin-wave excitations/magnon, introducing the metal atoms can enhance the $T_{\scriptsize{C}}$ up to the room temperature.   
   
\section{Conclusions}  
 We have performed the self-consistent non-collinear spiral calculations to investigate the effect of N$\acute{\textrm{e}}$el temperature $T_{\scriptsize{C}}$ in ZGNR under the electric field $E$ and the hole-electron doping $x$. We show that the $T_{\scriptsize{C}}$ cannot be increased by introducing $E$ and $x$. In addition, the reductions of $T_{\scriptsize{C}}$ are also followed by the reductions of spin stiffness $D$, thus there is a close relationship between the $T_{\scriptsize{C}}$ and $D$. These features are caused by the small magnetic moment of magnetic edge carbon atoms, making the magnetic instability.

We also show that the $T_{\scriptsize{C}}$ and $D$ reduce more rapidly in the large ribbon width $N$ than those in the small $N$ under $E$ but not under $x$. These features are caused by the exchange interaction $J_{ij}$ between two edge carbon atoms, i.e., the small ribbon width gets the largest $J_{ij}$. So, the large  $J_{ij}$ can compensate the rapid reductions of the $T_{\scriptsize{C}}$ and $D$ at the large $E$. On the contrary, no rapid reductions of $T_{\scriptsize{C}}$ and $D$ are observed as $x$ increases. Based on the results, introducing $E$ and $x$ never gives the $T_{\scriptsize{C}}$ close to the room temperature.
\section*{Acknowledgments}
A personal high computer has been used to performed the computations. We hereby state that this is an independent research without any fundings.

\end{document}